\def\draftversion{false}
\RequirePackage{ifthen}
\ifthenelse{\equal{\draftversion}{true}}{
  \documentclass[aps,prb,10pt,galley,amsmath,amssymb,showpacs,letterpaper]{revtex4}
}{
  \documentclass[aps,prb, 10pt,twocolumn,amsmath,amssymb,showpacs,letterpaper]{revtex4}
}

\usepackage{graphicx}
\usepackage{epsfig}
\usepackage{dcolumn}
\usepackage{bm}
\usepackage{subfigure}
\usepackage{hyperref}

\hypersetup{backref=true, pdfnewwindow=true, colorlinks=true,linkcolor=blue, anchorcolor=blue, citecolor=blue, filecolor=blue, menucolor=blue, urlcolor=blue}

\usepackage[usenames,dvipsnames]{color}

\ifthenelse{\equal{\draftversion}{true}}{
  \marginparwidth 2.7in
  \marginparsep 0.5in
  \newcounter{comm} % counter for commentaries
  \def\commnext{\stepcounter{comm}}
  \def\commtext{{\bf\color{blue}[\arabic{comm}]}}
  \def\commmar{{\bf\color{blue}[\arabic{comm}]}}
  \def\dvm#1{\commnext\marginpar{\small DV\commmar: #1}\commtext}
  \def\mtm#1{\commnext\marginpar{\small MT\commmar: #1}\commtext}
  \def\kgm#1{\commnext\marginpar{\small KG\commmar: #1}\commtext}
  \def\mlab#1{\marginpar{\small\bf #1}}
  
}{
  \def\dvm#1{}
  \def\mtm#1{}
  \def\kgm#1{}
  \def\mlab#1{}
  
}

\def\beq{\begin{equation}}
\def\eeq{\end{equation}}

\def\k{\textbf{k}}
\def\b{\textbf{b}}

\def\G{\textbf{G}}

\def\sg{{\cal S}}
\def\sgw{{\cal S}_{\rm W}}
\def\sgs{{\cal S}_{\rm S}}
\def\pg{{\cal G}}
\def\pgw{{\cal G}_{\rm W}}
\def\pgs{{\cal G}_{\rm S}}

\begin{document}
%============================================================

\title{Wannier Center Sheets in Topological Insulators}

\author{Maryam Taherinejad}
\email{mtaheri@physics.rutgers.edu}

\affiliation{Department of Physics and Astronomy, Rutgers University,
Piscataway, New Jersey 08854-0849, USA}

\author{Kevin F. Garrity}

\affiliation{Department of Physics and Astronomy, Rutgers University,
Piscataway, New Jersey 08854-0849, USA}

\author{David Vanderbilt}

\affiliation{Department of Physics and Astronomy, Rutgers University,
Piscataway, New Jersey 08854-0849, USA}

\date{\today}


\begin{abstract}
We argue that various kinds of topological insulators (TIs) can be
insightfully characterized by an inspection of the charge centers
of the hybrid Wannier functions, defined as the orbitals obtained
by carrying out a Wannier transform on the Bloch functions in
one dimension while leaving them Bloch-like in the other two.  From
this procedure, one can obtain the Wannier charge centers
(WCCs) and plot them in the two-dimensional projected Brillouin
zone.  We show that these WCC sheets contain the same kind of
topological information as is carried in the surface energy bands,
with the crucial advantage that the topological properties of the
bulk can be deduced from bulk calculations alone.  The distinct
topological behaviors of these WCC sheets in trivial, Chern,
weak, strong, and crystalline TIs are first illustrated by
calculating them for simple tight-binding models.  We then
present the results of first-principles calculations of the WCC
sheets in the trivial insulator Sb$_2$Se$_3$, the weak TI KHgSb,
and the strong TI Bi$_2$Se$_3$, confirming the ability of this
approach to distinguish between different topological behaviors
in an advantageous way.

\end{abstract}

\maketitle

%============================================================
\section{Introduction}
\label{sec:intro}
%============================================================

Since the work of Thouless {\it et al.}\cite{tknn-prl82} relating the Chern
number to the integer quantum Hall effect, there has been great
interest in insulators with topologically non-trivial band structures.
In time-reversal invariant insulators, the first Chern number vanishes,
but topologically non-trivial band structures can still emerge in
systems with strong spin-orbit coupling\cite{kane-prl05,kane2-prl05,
fu-prl07,moore-prb07} or crystal point group
symmetries.\cite{fu-prl11}  These topological phases are classified by
a series of $Z_2$ invariants.  In two dimensions, a single $Z_2$
invariant distinguishes a quantum spin Hall system from a trivial 2D
insulator, while in three dimensions, a total of four $Z_2$ invariants
$[\nu_0 ,\nu_1 ,\nu_2 ,\nu_3 ]$ are needed to classify the trivial,
weak, and strong topological phases which can emerge.
The topologically
non-trivial phases are gapped in the bulk, like trivial insulators,
but they are required to have robust metallic states on the edge (2D)
or surface (3D). These surface states provide the strongest
experimentally accessible signature of insulators with non-trivial
topology\cite{chen-sci09,
hsieh-nat08,xia-nat09,Analytis-nat09,Hsieh-prl09,
park-prb10,lin-nat10,hassan-rmp10}. However, for reasons of both
computational efficiency and theoretical clarity, it is preferable to
be able to calculate and understand the topological phases of insulators
purely from bulk calculations.

There have been several previously proposed methods for calculating
$Z_2$ invariants.  In principle, it is possible to calculate them by integrating
the Berry connection on half of the Brillouin
zone (BZ),\citep{fukui-jpsj07} but this
method requires fixing the gauge of the wavefunctions, which is
challenging in numerical calculations. In the special
case of a centrosymmetric crystal, the $Z_2$ invariants can be
calculated simply by considering the parity eigenvalues of the
occupied electronic states at the time-reversal-invariant (TRI)
momenta.  Our current work is closely related to a recently
developed method which is both general (not limited to crystals with
special symmetries) and computationally
efficient.\cite{soluyanov-prb11,yu-prb11}
This method relies
on the use of hybrid Wannier functions (WFs),
which provide an alternative
to the Bloch representation of the occupied band subspace.
By following the evolution of hybrid WFs
around a closed loop in the BZ, we can  describe the adiabatic, unitary
evolution of the occupied Bloch bands.  The partner switching of these
Wannier charge centers (WCC) around a closed loop, which describes a
pumping of ``time-reversal polarization,''
has been employed to calculate the $Z_2$
invariants in TRI insulators.\cite{fu-prb06,soluyanov-prb11,yu-prb11,Alexandradinata-arx12}

In this work, we focus on the topological
properties of WCCs in 3D materials, which
are functions of momentum $k$ in two
dimensions and can be plotted as sheets
over the 2D BZ.
We study the WCC sheets in trivial, Chern, weak
topological, strong topological,
and crystalline topological insulators (TIs) using tight-binding
models and first-principles calculations.
Although a knowledge of the
behavior of the WCCs on the TRI planes in the BZ is already sufficient for
determining the topological phase of the insulator, the more general behavior of
WCCs sheets in different topological phases, including the crystalline
topological phase, can provide new insights into the origin and properties
of these phases. In addition, unlike the surface states, the behavior of these
sheets is independent of surface termination and depends purely on the
bulk wavefunctions, allowing for a simpler picture of many properties.

The manuscript is organized as follows.  In Sec.~\ref{sec:wcc}
we define the WCC sheets, explain how to construct them,
and discuss their symmetry and topological properties.
In Sec.~\ref{sec:models} we introduce the tight-binding models
that will be used for illustrative calculations.  We also
present the materials systems that will be the subject of
first-principles calculations, and discuss the details of the
computational methods.  The calculated WCC sheets are presented
and discussed in Sec.~\ref{sec:results}, and we end with a summary
in Sec.~\ref{sec:sum}.

%============================================================
\section{WCC Sheets }
\label{sec:wcc}
%============================================================

The electronic ground state in periodic
crystalline solids is naturally described by
extended Bloch functions $\vert\psi_{n\mathbf{k}}\rangle$,
or the cell-periodic versions
$\vert u_{n\mathbf{k}}\rangle=e^{-i\bf k\cdot r}\vert
\psi_{n\mathbf{k}}\rangle$,
labeled by the band $n$ and crystal momentum
$\mathbf{k}$.
An alternative representation is the set of
localized orbitals or Wannier functions (WF)
which are defined in relation to the Bloch functions
by a unitary transformation:
\begin{equation}
\vert W_{n}(\mathbf{R})\rangle =
\dfrac{1}{(2\pi)^3}\int_{BZ} d\mathbf{k} \,
e^{i\bf{k}\cdot ({\mathbf{r}} -R)}
\vert {u_{n\mathbf{k}} \rangle}.
\label{eq:wf}
\end{equation}
These WFs are not unique, as the
$\textit{U}(N)$ gauge freedom in
choosing the $N$ representatives of the
occupied space at each $k$-point,
$\vert \tilde{u}_{n\mathbf{k}}
\rangle = \sum_m U_{mn}(\mathbf{k})
\vert {u}_{m\mathbf{k}} \rangle$, leaves
them gauge-dependent.

In 1D there is a unique gauge that
minimizes the spread functional of the
WFs.\cite{marzari-prb97}
These maximally localized WFs are
eigenfunctions of the band-projected
position operator $PzP$, where
$P=\sum_{nk} \vert\psi_{nk}\rangle
\langle\psi_{nk}\vert$ is the
projection operator onto the occupied
bands.
In 2D and 3D, on the other hand, the WFs
cannot be maximally localized in all directions
simultaneously, because the operators
$PxP$, $PyP$, and $PzP$ do not commute
and it is not possible to choose the WFs
to be simultaneous eigenfunctions of all
three. Instead, a compromise can be achieved
through an iterative procedure that localizes
the WFs in all directions as much as possible.\cite{marzari-prb97}

Insulators for which the occupied bands are characterized by
a nonzero Chern number are known as ``Chern'' or ``quantum
anomalous Hall'' (QAH) insulators.  In this case, it is
well-known that there is a topological
obstruction to the construction of exponentially
localized WFs.\cite{thouless-jpc84, thonhauser-prb06}
The vanishing of the Chern number in TRI insulators
guarantees the existence of localized WFs,
but special care needs to be taken in
choosing the gauge for $Z_2$-odd insulators,
as the localized WFs can only be constructed
in a gauge which does not let them come in
time-reversal pairs.\cite{soluyanov1-prb11}

The fact that there is never a topological obstruction to 
the construction of WFs in 1D suggests that a convenient
strategy for higher dimensions may be to construct ``hybrid
WFs'' that are Wannier-like in 1D and Bloch-like in the
remaining dimensions.\cite{sgiarovello-prb01,marzari-prb97}
Choosing the $\hat{z}$ direction for Wannierization in 3D,
these take the form
\begin{equation}
\vert W_{n l_z}(k_x,k_y)\rangle =
\dfrac{1}{2\pi}\int dk_z
e^{i\mathbf{k}\cdot ({\mathbf{r}} -l_z c
\hat{z})} \vert u_{n,\mathbf{k}} \rangle
\label{eq:hwf}
\end{equation}
where $l_z$ is a layer index
and $c$ is the lattice constant along $\hat{z}$.
Since there is a unique construction of maximally localized
WFs in 1D, these are easily constructed at each $(k_x,k_y)$,
regardless of whether the system is a normal insulator
or a Chern, $Z_2$, crystalline, or any other kind of
TI.
The charge center of these hybrid WFs
along the localized direction $\bar{z}_n$
is defined as the expectation value
$\bar{z}_n(k_x,k_y)=\langle W_{n0}
\vert \hat{z} \vert W_{n0} \rangle$
of the position operator $\hat{z}$
along this direction for the WF in the
home unit cell $\mathbf{R}=0$.
These WCCs, which are eigenvalues of
the $PzP$ operator, have been useful in
defining polarization
in 2D Chern insulators,\cite{coh-prl09}
understanding polarization in 3D layered
insulators,\cite{wu-prl06} and calculating
the $Z_2$ topological invariants in
TRI insulators.\cite{soluyanov-prb11}  Their sum
over occupied bands also gives the ``polarization 
structure'' describing the Berry-phase contribution
to the electeric polarization as a function of $\bf{k}$
in the 2D BZ.\cite{Yao-prb09}

It is well known that the nontrivial topology of Chern, $Z_2$,
and crystalline TIs is reflected in a
corresponding nontrivial connectivity of the surface energy bands.
While $k_z$ is clearly no longer a good quantum number
for a surface normal to $\hat{z}$,
$k_x$ and $k_y$ are still conserved,
so that if surface states appear in the bulk energy gap,
their energy dispersions $\epsilon_n(k_x,k_y)$
are good functions of momenta in the surface BZ.
In a similar way, the WCCs
$\bar{z}_n(k_x,k_y)$ can be plotted over
the same 2D BZ, where the Wannierized real-space
direction plays a role analogous
to the surface normal. Unlike the surface
states $\epsilon_n(k_x,k_y)$, the WCCs
$\bar{z}_n(k_x,k_y)$ depend only on
bulk properties. However, they still carry the same kind of
topological information as is contained in the surface states,
as will be explained in Sec.~\ref{sec:s&t}.

The WCCs can be obtained from a
parallel-transport-based construction\cite{marzari-prb97,wu-prl06}
in a straightforward way, as explained next.

%============================================================
%\subsection{CONSTRUCTION }
\subsection{Construction}
\label{sec:cons}
%============================================================

A cell-periodic Bloch state $\vert u_\k \rangle$ belonging
to an isolated band can be parallel transported to
$\vert u_{\k+\b}\rangle$ by choosing the phase of the latter such
that the overlap $\langle u_\k\vert u_{\k+\b} \rangle$
is real and positive, so that the change in the state is
orthogonal to the state itself.
If this is carried out repeatedly for a $k$-point string
extending along the $k_z$ direction by a
reciprocal lattice vector $\G\!\parallel\!\hat{z}$, then once the
phase of the initial $\vert u_{\k_0}\rangle$ is chosen, the
phase of each subsequent state, including the final
$\vert u_{\k_0+\G}\rangle$, is determined by
this parallel-transport procedure.  The phase of the last state
on the string is
then compared with the one $\vert \tilde{u}_{\k_0+\G}\rangle$
obtained by applying the periodic gauge condition
$\vert\psi_{\k_0+\G}\rangle=\vert\psi_{\k_0}\rangle$,
i.e., $\tilde{u}_{\k_0+\G}(\textbf{r})=\exp(-i\G\cdot\textbf{r})
\,u_{\k_0}(\textbf{r})$, and the phase mismatch
$U=\langle u_{\k_0+\G} \vert \tilde{u}_{\k_0+\G} \rangle$
is computed.  For a $k$-point string at $\k_\perp$ in the 2D BZ, this
yields the Berry phase $\phi(\k_\perp)=-\textrm{Im\,ln\,}U(\k_\perp)$ and
the WCC $\bar{z} (\k_\perp) = (c/2\pi)\,\phi(\k_\perp)$,
where $c$ is the lattice constant along $\hat{z}$.
If the parallel-transported states themselves are not needed, the
same result can be obtained more straightforwardly by computing
$\phi=-\textrm{Im\,ln\,}\prod \langle u_\k\vert u_{\k+\b} \rangle$,
where the product is carried out along the string and the
phases are chosen arbitrarily except for the periodic gauge condition
that fixes the phase of the first and last $k$-points in relation
to each other.

In the multiband case, where $n$ occupied bands are treated
as a group regardless of possible internal crossings or degeneracies,
the corresponding ``non-Abelian'' Berry phases $\phi_n$ can be
determined by generalizing this procedure in terms of
$n\times n$ matrix operations.  For each pair of neighboring
points along the string, the matrix $M_{mn}^{(\k,\k+\b)}=
\langle u_{m\k} \vert u_{n,\k+\b} \rangle$ is computed and
subjected to the singular value decomposition
$M=V\Sigma W^{\dagger }$, where $V$ and $W$ are unitary and
$\Sigma$ is real and diagonal (typically, nearly unity).
Again, the states at the end point $\k_0+\G$ are predetermined by
those at the start $\k_0$ by the periodic gauge condition.
We can then identify $U^{(\k,\k+\b)}=V W^{\dagger }$ as the
unitary rotation from $\k$ to $\k+\b$, and the global unitary
rotation matrix $\Lambda(\k_\perp)=\prod U^{(\k,\k+\b)}$ is
constructed as the product of these along the string.
Being unitary, its eigenvalues $\lambda_n$ are unimodular,
and we can identify the non-Abelian Berry phases (also known
as Wilson loop eigenvalues) as
$\phi_n(\k_\perp)=-\textrm{Im\,ln\,}\lambda_n(\k_\perp)$.
The WCCs are then just
\begin{equation}
\bar{z}_n(\k_\perp)=\frac{c}{2\pi}\,\phi_n(\k_\perp) \,.
\label{eq:wcc}
\end{equation}
As discussed in Ref.~\onlinecite{marzari-prb97}, this procedure
gives the centers of the maximally-localized Wannier functions in 1D
algebraically, without the need for any iterative
localization procedure; we just repeat this procedure for each
$\k_\perp$ to construct the WCC sheets.

%============================================================
\subsection{SYMMETRIES AND TOPOLOGY}
\label{sec:s&t}
%============================================================

A major theme of the present work is to show how the
WCC sheet structure $\bar{z}_n (\k_\perp)$ shares many
qualitative features with the surface energy bandstructure
$\epsilon_n(\k_\perp)$, with the crucial advantage that they
can be used to deduce the topological properties of the
bulk from bulk properties alone.  In this subsection we
show that the WCC sheet structure obeys all of the symmetries
that are found in the surface energy bandstructure, and
sometimes more.  In particular, when TR is present, the Kramers
degeneracies found at the 2D time-reversal invariant momenta (TRIM)
in the surface energy band
structure also necessarily appear in the WCC sheet structure.
We also sketch a physical argument as to why the topological
connectedness of the WCC sheets mirrors that of the surface
bandstructure, providing access to the topological indices in
a similar way.

\subsubsection{Symmetry}

To review, we consider a crystalline insulator
with $\hat{z}$ taken along a primitive reciprocal
lattice vector, and let $k_\parallel$ and $\k_\perp$ denote
the wavevectors parallel and perpendicular to
$\hat{z}$ respectively.  We then consider the surface
bandstructure $\epsilon_n(\k_\perp)$ for a 1$\times$1 (unreconstructed)
surface that has been cut normal to $\hat{z}$, where $n$ labels
energy eigenstates lying in the bulk projected band gap.  We also
consider the WCC sheets $\bar{z}_n (\k_\perp)$ constructed
as detailed in Sec.~\ref{sec:cons}, where $n$ labels the
sheets with $-c/2\le z\le c/2$ in one unit cell along $z$. In both cases,
$\k_\perp$ resides in the same 2D surface BZ (both functions
have the same periodicity in $\k_\perp$).

An element $S=\{G|\bm{\tau}\}$ of the full space group $\sg$
is composed of a generalized rotation $G$ (possibly improper,
and possibly containing TR) followed by a possible fractional
translation $\bm{\tau}$ (in non-symmorphic crystals), in addition
to lattice translations; the full point group
$\pg$ is composed of all of the $G$ appearing in the space-group
elements.

The symmetry of the WCC sheets is controlled by the reduced space
group $\sgw\subseteq\sg$ and the corresponding point
group $\pgw\subseteq\pg$ defined by restricting the list of $G$'s
to those that map $\hat{z}$ onto $\pm\hat{z}$.  For such operations,
let $G=K T_z G_\perp$ where $G_\perp$ is the in-plane
rotation (possibly improper),
$T_z$ is either the identity or the simple mirror $M_z$,
and $K$ is either the identity or TR.
Then a space-group element $\{G|\bm{\tau}\}\in\sgw$ must
transform a hybrid Wannier function $W_n(\k_\perp)$
into another hybrid Wannier function $W_{n'}(\pm G\k_\perp)$,
with the Wannier center transformed as
\beq
\bar{z}_{n'}(\pm G_\perp\k_\perp)=T_z\bar{z}_n(\k_\perp)+\tau_z  \,,
\label{eq:wccsym}
\eeq
where the minus sign applies if $G$ contains TR.

The symmetry of
the surface bandstructure $\epsilon_n(\k_\perp)$, on the other
hand, is associated with the space group $\sgs\subseteq\sgw$ with the additional
constraints that its elements do not reverse $\hat{z}$
to $-\hat{z}$ and and do not contain partial translations $\tau_z$
along $\hat{z}$.  Then for any element $G=KG_\perp$ in the corresponding
point group $\pgs$ we have that
\beq
\epsilon_n(\pm G_\perp\k_\perp)=\epsilon_n(\k_\perp)
\label{eq:surfsym}
\eeq
where again the minus sign applies if TR is involved.

Since $\pgs\subseteq\pgw$, it follows from Eq.~(\ref{eq:wccsym})
that the WCC sheets also obey
\beq
\bar{z}_n(\pm G_\perp\k_\perp)=\bar{z}_n(\k_\perp)
\label{eq:wccweaksym}
\eeq
for any $G\in\pgs$.  Thus, the WCC sheets show at least as much
symmetry as the surface bandstructure.  If the space group
contains symmetry elements that reverse the $z$ axis, then there is
an additional symmetry $\bar{z}_n(\pm G_\perp\k_\perp)=-\bar{z}_n(\k_\perp)$
associated with these elements, or if it contains
glide or screw operations along $\hat{z}$, then also
$\bar{z}_n(\pm G_\perp\k_\perp)=\bar{z}_n(\k_\perp)+\tau_z$.  These
additional WCC symmetries have no counterpart in the surface
bandstructure.

Finally, we note that TR symmetry plays a similar role
for the WCC sheets as for the surface energy bandstructure.
Specifically, for any $G\in\pgs$, a Kramers degeneracy is enforced whenever
$G_\perp\k_\perp=-\k_\perp$ (modulo a reciprocal lattice vector),
due to the antiunitary nature of the TR operation.  In particular,
if TR by itself is a symmetry, then the WCC sheets and the surface
energy bands are guaranteed to touch and form Kramers pairs at
all of the TRIM.  Additionally, if $C_2^Z \otimes \hbox{TR}$
is a symmetry, then both the WCCs and surface energy bands are
Kramers degerate everywhere in the 2D BZ.

\subsubsection{Topology}

Just as the symmetries of the surface bandstructure are
replicated in the WCC sheet structure, a similar principle
applies to the topological properties.  This will be amply
illustrated by the examples to follow, but we give here a
sketch of a general argument that this should be so.

%---------------------------------------------
\begin{figure}
\includegraphics[width=0.85\columnwidth]{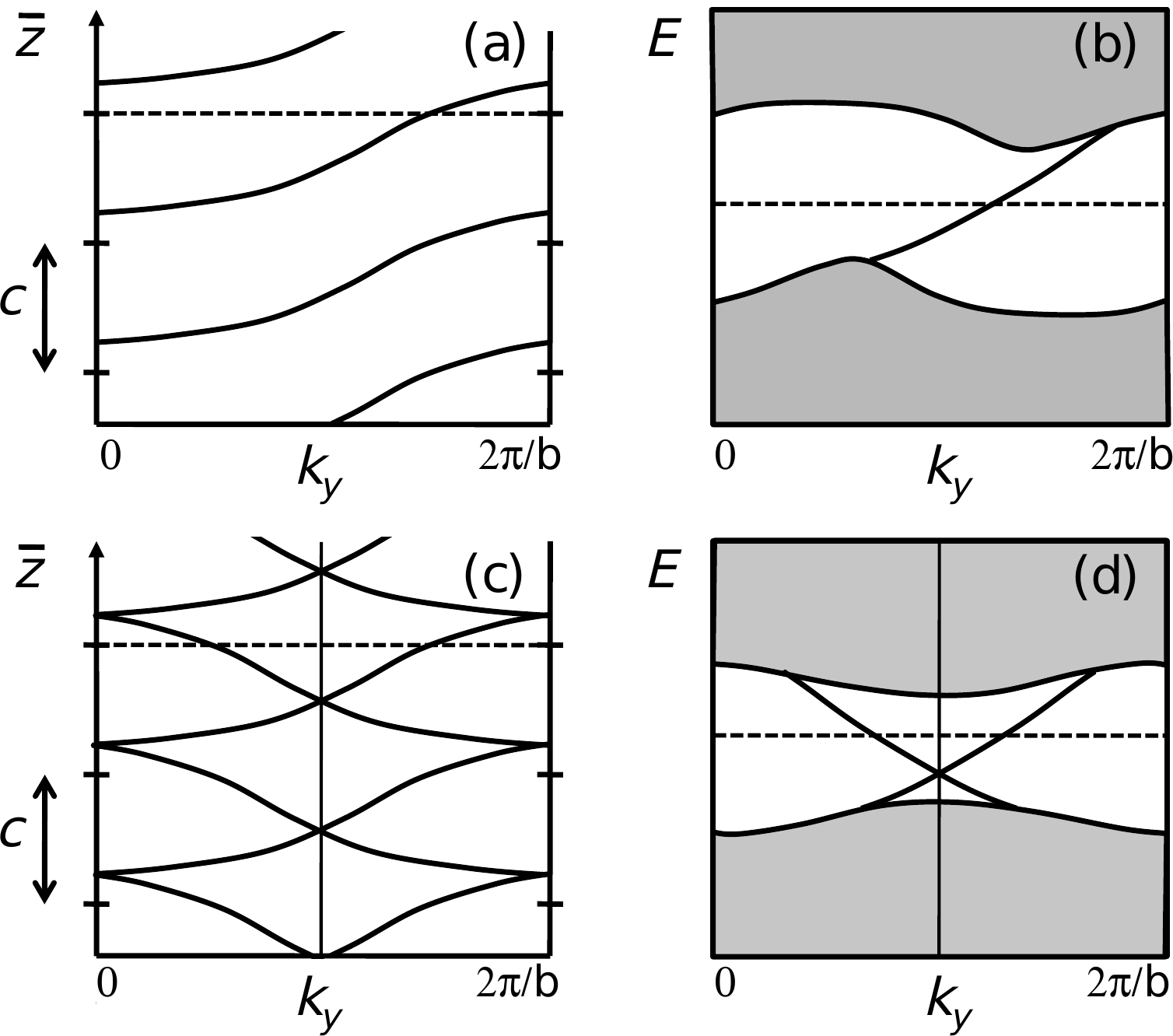}
\caption{\label{fig:bzpump}
(a) Flow of Wannier charge centers along $\hat{z}$ vs.\ $k_y$
for a 2D Chern insulator.
(b) Flow of surface energy bands vs.\ $k_y$
for a 2D Chern insulator.
(c-d) Same, but for a 2D $Z_2$-odd (quantum spin Hall) insulator.
Dashed lines are arbitrary reference positions in (a) and (c), or
Fermi energies in (b) and (d).}
\end{figure}
%---------------------------------------------
For simplicity, consider first a 2D Chern insulator lying in
the $y$-$z$ plane with one occupied band carrying a Chern number
$C=+1$.  Then the WCC $\bar{z}(k_y)$ undergoes a shift by
$c$ as $k_y$ is adiabatically carried from $k_y=0$ to $k_y=2\pi/b$
(assuming a rectangular $b\times c$ unit cell), as shown in
Fig.~\ref{fig:bzpump}(a).  This means that
one electron is adiabatically pumped by $c\hat{z}$ during
one cycle of $k_y$ around the 1D BZ.  If the edge
bandstructure remained gapped throughout the cycle, this would
lead to a contradiction, since by conservation of charge one
extra electron per surface unit cell would reside on the
top edge at the end of the cycle.  However, the starting and
ending point are physically identical,
so the edge charge must be the same.  This paradox can
only be avoided if there is a surface state that emerges from
the valence band, rises throught the gap, and disappears into
the conduction band during one cycle, as shown in Fig.~\ref{fig:bzpump}(b).
In this case, the sudden
loss of one electron that occurs when the surface band crosses
the Fermi energy compensates for the gradual gain of one electron
from the pumping, restoring charge conservation.  In other words,
we conclude that the edge bandstructure has a state crossing the
gap if and only if the WCC structure has a WCC that winds by
one unit during the cycle.

More generally, for an insulator with
$N$ occupied bands, if $\sum_n^N \bar{z}_n/c$ winds by Chern
integer $C$ during the cycle, the number of up-crossing minus the
number of down-crossing surface bands in the edge bandstructure
must equal $C$ in order to satisfy charge conservation.
The argument also generalizes to 3D Chern insulators.  If the
WCC sheet structure is computed in the $\hat{z}$ direction,
the Chern indices along $x$ and $y$ (that is, corresponding
to Berry curvatures $\Omega_{yz}$ and $-\Omega_{xz}$) are
evident in the $z$-windings of the WCC sheets as $\k_\perp$ is
cycled in the $k_y$ and $k_x$ directions respectively.  In
each case, a similar surface state crossing necessarily must also
occur in the surface bandstructure, following the arguments given
above.

Turning now to TR-invariant insulators, the Chern number always
vanishes, being odd under TR-symmetry, but the WCC structure
and surface band structure still share their
topological properties.  Recall that
TR symmetry leads to double degeneracy in both the
the WCCs and the surface energy bands at TRI points in BZ. 
Here we show that the WCCs connect the
TRI points in the same manner as the energy bands do,
and can be used in a similar way to deduce the $Z_2$ index of
the system.

Consider the simple case of a 2D $Z_2$ insulator in the
$y$-$z$ plane with two occupied bands. The TR symmetry relates
the WCCs and the surface energy bands in the second half of
the BZ to the ones in the first half by $z(k_y)=z(2\pi/b-k_y)$
and $\epsilon(k_y)=\epsilon(2\pi/b-k_y)$, as illustrated in
Figs.~\ref{fig:bzpump}(c-d), so we only need to study their
behavior in the first half $[0,\pi / b]$.  In the absence of
spin-mixing terms,
the system decouples into two independent
insulators with equal and opposite Chern numbers for spin-up
and spin-down electrons.
If these are $\pm1$, the system is $Z_2$-odd.  This implies both
that the WCCs must switch partners as $k_y$ evolves from 0 to
$\pi/c$, as shown in Fig.~\ref{fig:bzpump}(c), and that the surface
energy bands zigzag, as shown in Fig.~\ref{fig:bzpump}(d).  More
precisely, an arbitrary horizontal line in Fig.~\ref{fig:bzpump}(c)
intersects the WCC curves just once (or, in general, an odd number
of times) in the half-BZ, as does an arbitrary Fermi level for the
surface energy bands in the half-BZ in Fig.~\ref{fig:bzpump}(d).
One unit of up spin, relative to down spin, is pumped to the
edge during this half-BZ evolution, corresponding to the ``TR
polarization pumping'' discussed by Fu and Kane.\citep{fu-prb06}
For a $Z_2$-even insulator, the number of crossings is,
instead, an even integer (typically zero) for both the WCCs
and the surface energy bands.

In more realistic $Z_2$ insulators, the spin-orbit interaction
mixes up and down spins such that the energy bands are no longer
perfectly spin-polarized and a spin-Chern classification of
the system is no longer guaranteed. However, as long as the
bulk energy gap remains open as these spin-mixing terms are
adiabatically turned on, neither the even/oddness of the
number of WCC crossings, nor the even/oddness of the number
of surface energy band crossings, can change.  Therefore,
it remains true that the $Z_2$ index deduced from the WCC
evolution is the same as that deduced from the surface energy
bands, i.e., they both contain the same topological information.

The weak and strong topological indices of a 3D TR-invariant
insulator can be determined from the 2D indices on the six
TRI faces $\tilde{k}_j\!=\!\{0,\pi\}$ of the 3D BZ
(where $\tilde{k}_x\!=\!k_xa$,
$\tilde{k}_y\!=\!k_yb$, $\tilde{k}_z\!=\!k_zc$),
which are negative if the WCCs have a non-trivial
connectedness on that face and positive otherwise.
Assigning an index $\nu(\tilde{k}_i)$ to each of these faces, the
four $Z_2$ invariants $[\nu_0 ,\nu_1 ,\nu_2 ,\nu_3 ]$ that
uniquely specify the topological phase of a TR-invariant
insulator can be determined from
these $\nu(\tilde{k}_i)$ as follows. The three
$\nu_i\equiv\nu(\tilde{k}_i=\pi)$,
which are known as weak topological indexes, are determined
from the WCC behavior on the $\tilde{k}_i=\pi$ faces, while the strong
topological index $\nu_0\equiv\nu(\tilde{k}_i=0)\nu(\tilde{k}_i=\pi)$ is only
negative if the topological indices of opposing TRI faces are
opposite, i.e., if the WCCs on the $\tilde{k}_i=0$ and
$\tilde{k}_i=\pi$ faces have different behavior.
The indices could similarly be deduced from the behavior
of the surface energy bands.
For both the WCC
and surface problems, we have to choose a particular axis
$\hat{z}$ to define $\bar{z}$ or as the surface normal, and
in this case we are only sensitive to four of the six TRI-face
indices, defining whether WCCs (or surface states) zigzag
or not along the four edges of the quarter 2D BZ.  This
determines the strong index $\nu_0$ and two of the three weak
indices ($\nu_1$ and $\nu_2$); the procedure has to be repeated
with a different choice of axis to obtain the third weak index
$\nu_3$.

In summary, we expect that the flow and connectedness of the WCC
sheets and the surface energy bands should always show the same
qualitative features.  Not surprisingly, similar considerations
apply to the case of crystalline TIs as well.
Numerous examples will be presented below which amply illustrate
this principle.

%============================================================
\section{MODELS AND CALCULATIONS}
\label{sec:models}
%============================================================

We study the properties of WCCs in different topological
phases using simple tight-binding (TB) models as well as realistic
density-functional theory (DFT) descriptions of known materials.
In particular, we use a Haldane-like\cite{Haldane-prl88}
TB model of spinless electrons on a hexa\-gonal lattice
to study the properties of the WCC sheets in a 3D Chern insulator;
the model of Fu, Kane, and Mele (FKM)\citep{fu-prl07} to study
the WCCs of trivial, weak, and strong topological phases; and the
the tetragonal TB model of Fu\citep{fu-prl11} to study
a crystalline TI.  These TB models
are described in Sec.~\ref{sec:tb models}.  We then compute the
behavior of the WCC sheets in the $Z_2$-even Sb$_2$Se$_3$, weak
$Z_2$-odd KHgSb, and strong $Z_2$-odd Bi$_2$Se$_3$ insulators
using first-principles DFT calculations.  These materials and their
crystal structure are described in Sec.~\ref{sec:r-material},
and the details of our computational approach are presented in
Sec.~\ref{sec:1st-p}.

%============================================================
\subsection{TIGHT-BINDING MODELS}
\label{sec:tb models}
%============================================================

A TB model of a 2D Chern insulator
was first introduced by Haldane on a honeycomb
lattice.~\citep{Haldane-prl88}
This spinless model is constructed by starting with real
first and second-neighbor hoppings,
but the time-reversal symmetry is then broken by introducing local
magnetic fluxes in a pattern that respects the symmetry of the
lattice and sums to zero in each unit cell. This magnetic flux
has the effect of multiplying the second-neighbor hoppings by
a unimodular phase factor $\lambda=e^{i\varphi}$. We then stack
these 2D layers in the normal direction to make a 3D TB model of
a Chern insulator:
\def\phdag{^{\phantom{\dagger}}}
\begin{eqnarray}
H&=&t_1\sum_{l,<ij>}c_{il}^\dagger c_{jl}\phdag
+t_2\sum_{l,\ll ij \gg} \lambda c_{il}^\dagger c_{jl}\phdag
\nonumber\\
&+t_1^{\prime}&\sum_{li}c_{il}^\dagger c_{i,l+1}\phdag
+t_2^{\prime}\sum_{l,<ij>}c_{il}^\dagger c_{j,l+1}\phdag + \hbox{H.c.}
\label{eq:tbmod}
\end{eqnarray}
Here $l$ is the layer index, single and double brackets 
label first- and second-neighbor in-plane pairs
with hoppings $t_1$ and $t_2$ respectively,
and $t_1^{\prime}$ and $t_2^{\prime}$ are (real) vertical
and nearest-diagonal interlayer hoppings respectively.
The hoppings included explicitly in the second term of
Eq.~(\ref{eq:tbmod}) are those for clock-wise hoppings around
the hexagon; counterclockwise ones are accounted for by
the Hermitian conjugation and have phases $\lambda^*$.
With $t_1=-1.0$, $t_1^{\prime}=-0.45$,
$t_2=0.15$, $t_2^{\prime}=0.015$, and $\varphi=0.5\pi$
the occupied band has a Chern number of one.

The FKM model\citep{fu-prl07} is a four-band TB model 
of $s$ states on a diamond lattice in 3D with a 
spin-orbit interaction, and takes the form

\begin{equation}
H=t\sum_{<ij>}c_i^\dagger c_j + i
(8 \lambda_{\rm so} / a^2)
\sum_{\ll ij \gg} c_i^\dagger \textbf{s} \cdot
( \textbf{d}_{ij}^1 \times \textbf{d}_{ij}^2) c_j .
\label{eq:3d-km}
\end{equation}
Here the first and second terms describe spin-independent
first-neighbor and spin-dependent second-neighbor hoppings
respectively; $\lambda_{\rm so}$ is the spin-orbit strength,
and $a$ is the cubic lattice constant, which is set to one.
The second-neighbor hopping between sites $i$ and $j$ depends on
spin and on the unit vectors $\textbf{d}_{ij}^{1,2}$ describing
the two first-neighbor bonds that make up the second-neighbor
hop.  For $t=1$ and $\lambda_{\rm so}=0.125$, the model 
has a gap closure at the high symmetry X point in the Brillouin
zone.

%--------------------------------------------------
\begin{table}
\caption{\label{tab:phase}%
Topological phase of the FKM model\citep{fu-prl07} as a
function of parameter $\alpha$ specifying the relative
strength of the [111] bond according to
$t_{111}=t(1+\alpha)$.}
\begin{ruledtabular}
\begin{tabular}{lll}
\mbox{$\alpha$}&\mbox{$[\nu_0;\nu_1 \nu_2 \nu_3]$}
&\mbox{Topological phase}\\
\hline
$(-\infty,-4)$&$[+;+++]$&Trivial insulator\\
\mbox{$(-4,-2)$}&$[-;---]$&Strong topological insulator\\
\mbox{$(-2, 0)$}&$[+;++-]$&Weak  topological insulator\\
\mbox{$(0,2)$}&$[-;++-]$&Strong  topological insulator\\
\mbox{$(2, \infty)$}&$[+;+++]$&Trivial insulator\\
\end{tabular}
\end{ruledtabular}
\end{table}
%-----------------------------------------------------------

By varying the relative strength of the nearest-neighbor
bond in the [111] direction, $t_{111}=t(1+\alpha)$,
the cubic symmetry is broken and the system can be switched
between trivial, weak and strong topological phases, as shown in
Table \ref{tab:phase}.  These insulating phases are separated
from each other by gap closures at $\alpha\!=\!-4$,
$-$2, 0, and 2.  For $\alpha\!<\!-4$ and $\alpha\!>\!2$,
the $t_{111}$ bond is stronger than the other bonds
and the system can be adiabatically transformed to
a system of dimers, which is toplogically equivalent to a trivial
atomic insulator. For $-2\!<\!\alpha<0$, on the other hand, the
$t_{111}$ bond is weaker than the others, and
the system can be considered as a collection of 2D spin-Hall layers
stacked along the [111] direction. Thus, the system is a weak
TI in this range of $\alpha$.
For $-4\!<\!\alpha<-2$ and
$0\!<\!\alpha\!<\!2$, $t_{111}$ is stronger than the other first-neighbor
bonds, but not strong enough to push the system into
the topologically trivial phase. As a result, the 3D KM model is a
strong $Z_2$ TI for these values of $\alpha$.

For studying the WCC sheet behavior in a topological
crystalline insulator, we adopted the TB model of Fu,\citep{fu-prl11}
consisting of a tetragonal lattice with two inequivalent A and B atoms
stacked above one another, each carrying $p_x$ and $p_y$ orbitals,
forming bilayers that we index by $n$.
The total system Hamiltonian can be written as

\begin{equation}
H = \sum_{n} \left( H_n^A+H_n^B+H_n^{AB} \right),
\label{eq:Hsum}
\end{equation}
where $H^A$ and $H^B$ are the contributions describing
intralayer hoppings while $H^{AB}$ describes interlayer ones.
The former are given by
\begin{equation}
H_n^X =\sum_{ij}t^X(\mathbf{r}_i-\mathbf{r}_j)
\sum_{\alpha, \beta}c^\dagger_{X\alpha}
(\mathbf{r}_i,n)e^{ij}_\alpha
e^{ij}_\beta c_{X\beta}(\mathbf{r}_i,n)
\label{eq:HAorB}
\end{equation}
and the latter by
\begin{eqnarray}
H_n^{AB} &=& \sum_{ij}t'(\mathbf{r}_i-\mathbf{r}_j)
\sum_{\alpha, \beta}\left[c^\dagger_{A\alpha}
(\mathbf{r}_i,n)c_{B\alpha}(\mathbf{r}_i,n)
+\hbox{H.c.}\right]\nonumber\\
&+&t'_z\sum_i\sum_\alpha\left[c^\dagger_{A\alpha}
(\mathbf{r}_i,n)c_{B\alpha}(\mathbf{r}_i,n_1)
+\hbox{H.c.}\right] .
\label{eq:HAB}
\end{eqnarray}
Here $\mathbf{r}\!=\!(x,y)$ labels the coordinate in the plane,
$X\!=\!\{A,B\} $ labels the sublattice, $\alpha$ and $\beta$
label the $\{p_x,p_y\}$ orbitals, and
$e^{ij}_\alpha$ is cosine of the angle between the bond
$(\mathbf{r}_i -\mathbf{r}_j)$ and orbital $p_{\alpha}$.
We choose the nearest- and
next-nearest-neighbor hopping amplitudes to be
$t^A_1=-t^B_1=1$ and $t^A_2=-t^B_2=0.5$ in $H^A$ and $H^B$, and $t'_z=2$,
$t'_1=2.5$ and $t'_2=0.5$ in $H^{AB}$.

Note that this TB model is spinless, as the
spin-orbit coupling plays no role in the non-trivial topology of
crystalline TIs.  Instead, the topological
classification is based on certain crystal point-group symmetries and TRI,
leading to robust surface states on those surface that respect
the symmetries in question. In the tetragonal Fu model, these
topological surface states exist on the (001) surface, where
the fourfold $C_4^z$ rotational symmetry of the crystal is
preserved.

%============================================================
\subsection{MATERIAL SYSTEMS}
\label{sec:r-material}
%============================================================

We carry out first-principles calculations of the WCC sheet structure
for Sb$_2$Se$_3$, KHgSb, and Bi$_2$Se$_3$ as prototypical realizations
of trivial, weak, and strong topological phases, respectively.
Bi$_2$Se$_3$ has
a rhombohedral layered structure with space group $D^5_{3d}
(R\bar{3}m)$. It consists of quintuple layers (QLs) formed by
stacking Se and Bi triangular-lattice planes in the order
Se-Bi-Se-Bi-Se, with two
identical Bi atoms, two identical Se atoms and a third Se atom at
the center.  These QLs have strong internal covalent bonding,
but the interaction between QLs is much weaker, being
largely of van der Waals type.
The states near the Fermi energy come from the

Bi $6p$ and Se $4p$ orbitals.  The strong SOC leads
to a band inversion at the $\Gamma$ point and makes this material a
strong $Z_2$ insulator with a band gap of 0.3\,eV.\cite{xia-nat09,Zhang-apl09}
Sb$_2$Se$_3$ shares the same
rhombohedral layered structure
as Bi$_2$Se$_3$, but the weaker
SOC in this material leaves it in
a topological trivial phase.

KHgSb consists of layers of HgSb in a honeycomb lattice, with hexagonal
layers of K atoms stuffed between them. In a single layer of KHgSb, the
valence bands near the Fermi energy are composed of the Hg $6s$ and Sb $5s$ and $5p$
states, while the K $4s$
band is considerably higher in energy.  The
strong SOC in the honeycomb HgSb layer leads to a band inversion at
the $\Gamma$ point
in the 2D BZ and makes an isolated KHgSb layer a 2D
TI. These 2D TI layers can be stacked along the
$z$ direction to form a 3D
lattice, but the inter-layer coupling is very weak and there is
little dispersion along the $\Gamma$-$Z$ direction.
These honeycomb layers can either be stacked in
an AA sequence to make a ``single-layer'' form, or in an ABAB sequence
to make a ``double-layer'' form, where B is rotated by 60$^\circ$ with respect to A. In the latter structure,
which is experimentally observed, the primitive cell contains two
honeycomb layers. Thus, two band inversions occur and cancel
each other out at
$\Gamma$, and the same happens at $Z$, making the compound a trivial
insulator.\citep{zhang-prl11,yan-prl12}
In the hypothetical single-layered structure,
which is proposed as an example of a weak TI,\citep{yan-prl12} there is
only one honeycomb layer in the primitive cell, and
a single band inversion happens at $\Gamma$ and another at $Z$.
Thus single-layered
KHgSb can be viewed as a stack of weakly coupled 2D TIs and belongs to
the weak $Z_2$-odd topological class.
Here, we focus on single-layered KHgSb, and we compare its WCC
sheets to the weak topological phase of the FKM TB model in
Sec.~\ref{sec:wz2odd}.

%============================================================
\subsection{COMPUTATIONAL METHODS }
\label{sec:1st-p}
%============================================================

Our first-principles calculations of
WCC sheets are based on DFT calculations
using the PBE exchange-correlation functional\citep{perdew-prl96}
performed with the Quantum Espresso package.\cite{quantum-espresso}
We use fully-relativistic optimized norm-conserving
pseudopotentials from the Opium package, with the semicore
Bi $5d$, Sb $4d$, Hg $5d$, and K $3s3p$ states included in
the valence. The self-consistent calculations
are carried out for the experimental structures using a
$10\times 10 \times 10$ Monkhorst-Pack\cite{monkhorst-prb76} $k$-mesh.
The plane-wave energy cutoff is set to 70\,Ry.

In principle one could include all occupied bands in the WCC
construction.  However, taking Bi$_2$Se$_3$ as an example,
the occupied Bi $5d$ semicore states and the shallow Bi $6s$ and
Se $4s$ bands
have an obvious atomic character and remain well separated
from the active valence $p$ bands, so they are clearly trivial
and do not need to be included in the topological analysis.
Therefore, we concentrate on constructing WCC sheets only for the
remaining upper valence bands.  As these are the lowest 18 of the
30 bands of Bi $6p$ and Se $4p$ character, we do this by constructing
a Wannier representation in this 30-band space using
the Wannier90 package\citep{mostofi-cpc08}
to generate an {\it ab initio} TB Hamiltonian from 
the DFT calculation. The frozen window in which the first-principles
band structure is exactly
reproduced extends from 2\,eV below to 2\,eV above the Fermi level
$E_F$.  From
the outer energy window, which extends to 20\,eV above $E_F$,
80 Bloch bands
are used to produce 30 WFs for the Bi, Sb, and Se 
$p$ bands in  Bi$_2$Se$_3$ and  Sb$_2$Se$_3$.
The orbital positions and hopping
parameters between them are then used to construct the effective
tight-binding Hamiltonians.
Similarly, for KHgSb we carry out the Wannier construction
for the ten Bloch bands of K 4$s$, Hg $6s$ and Sb $5p$ character,
of which the bottom six are the highest valence states. The outer window is
chosen at 14\,eV above $E_F$ for KHgSb, with ten WFs constructed from 20 Bloch
bands.

We have implemented the calculation of the WCC sheets into
Version 1.6.2 of the open-source PythTB tight-binding code
package.\citep{pythtb} The Wannierized Hamiltonians are imported
into the PythTB code to calculate the WCC sheets using the 
parallel-transport approach explained in Sec.~\ref{sec:cons}.

%============================================================
\section{RESULTS}
\label{sec:results}
%============================================================

In this section, we present the WCC sheets for the different topological
phases we have studied. 
For the 3D Chern insulator in Sec.~\ref{sec:chern},
the WCC sheets are plotted over the entire 2D BZ,
while for the TR-invariant systems of Secs.~\ref{sec:z2even}-\ref{sec:sz2odd}
the sheets are plotted over one quarter of the BZ, i.e., between
the TRI momenta $[0,0]$, $[0,\pi]$, $[\pi,\pi]$,
and $[\pi,0]$.

The axis of highest rotational symmetry in each TB model or
material system is chosen as the $z$ axis.
This axis in the FKM model is along the bond with altered strength
$(t_{111})$; the model has a 3-fold symmetry around this axis,
which when combined with TR-symmetry results in a 6-fold rotational
symmetry in the 2D BZ.
In Sb$_2$Se$_3$ and Bi$_2$Se$_3$ the $z$ axis
is normal to the quintuple layers, which is
the axis of 3-fold symmetry. In KHgSb the
$z$-axis is chosen normal to the honeycomb HgSb
layers, and in the Fu tetragonal TB model
it is along the tetragonal axis.

The WCC sheets are computed along both $z$ and
$y$ and plotted
versus $(k_x,k_y)$ and  $(k_x,k_z)$ respectively.
(Henceforth we shall not be careful about the distinction
between $k_x$ and $\tilde{k}_x=k_xa$, etc.; the meaning should
be clear from the context.)
Plotting the WCC sheets along these two perpendicular
directions is especially important to
reveal the topological behavior in the 3D Chern, weak $Z_2$,
and topological crystalline  phases,
where, as we shall see, the topology
of the WCC sheets may look trivial in one direction
but topological in another.

The WCC sheets for the TR-broken Chern insulator phase are
discussed next.  WCC sheets for the TR-invariant trivial, weak, and strong
$Z_2$ phases are discussed in
Secs.~\ref{sec:z2even}-\ref{sec:sz2odd}, using the FKM
model and its material system analogues in each
phase. The WCC sheets for the crystalline topological
phase are discussed in Sec.~\ref{sec:crystalline}.

%============================================================
\subsection{TR-broken Chern insulator}

\label{sec:chern}
%============================================================

%-------------------------------------------
\begin{figure}
\includegraphics[width=1.\columnwidth]{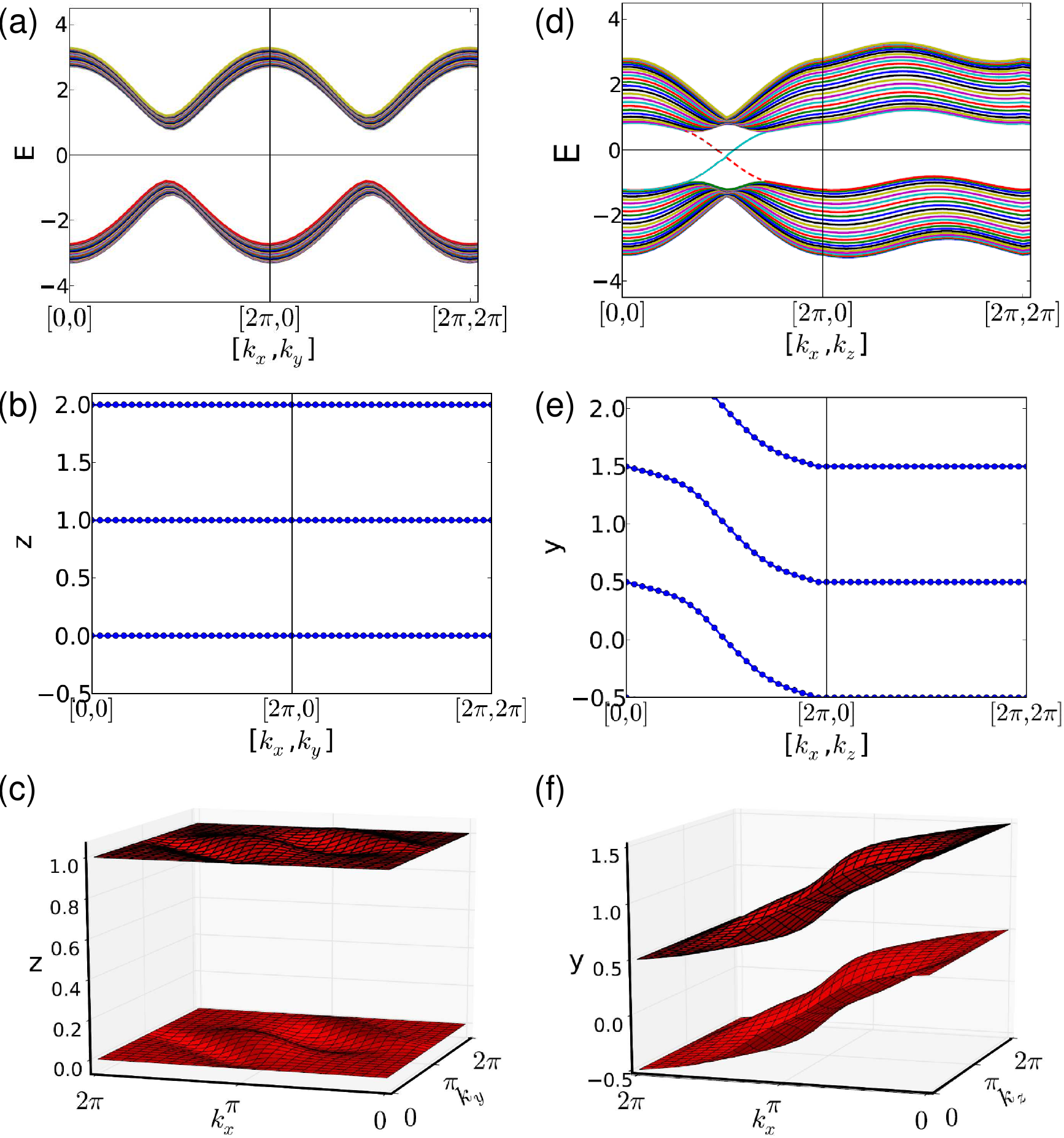}
\caption{\label{fig:chern}
(Color online)
Surface energy bands and WCC sheets for the TR-broken Chern insulator model.
(a-c) Surface normal and WCCs along $\hat{z}$ vs.\ $(k_x,k_y)$.
(d-f) Surface normal and WCCs along $\hat{y}$ vs.\ $(k_x,k_z)$.
Surface states for a 24-layer slab in (a) and (d);
WCCs around 2D BZ boundary in (b) and (e);
WCCs in 2D BZ in (c) and (f).
Dashed and solid surface states in (d) reside on the top and
bottom of the (010) slab respectively.
The WCC sheets and surface bands wind by one unit in the $k_y$-$k_z$
plane, but not in the $k_x$-$k_y$ plane.}
\end{figure}
%---------------------------------------------

We first consider the TB model for a TR-broken Chern
insulator phase that was introduced in Sec.~\ref{sec:tb models}.
It is composed of 2D Chern layers stacked along the $z$
direction with weak interlayer coupling, so we do not expect
an (001) slab of the 3D model to show any topological surface
states.  This is confirmed in the surface projected bandstructure
plotted in Fig.~\ref{fig:chern}(a), where the shaded
region indicates the region of bulk energy bands.  No
surface states are visible in this case, consistent with
the trivial topology for this orientation.
By the same token, the WCC sheets computed along the $z$
direction from the single occupied band remain localized in
the vicinity of the $z$ positions of the layers, with
no topological evolution along $k_x$ or $k_y$.
This is shown in Fig.~\ref{fig:chern}(b-c), where the
WCC sheets are plotted around the boundary, and throughout
the interior, of the 2D projected BZ respectively.

In contrast, any slab of 
the 3D system that cuts through the 2D Chern layers 
will reveal the topological nature of the 3D crystal by
displaying a surface energy band traversing the bulk gap
on each surface, as shown in Fig.~\ref{fig:chern}(d) for
a (010) slab.  The corresponding
$\bar{y}(k_x,k_z)$ WCC sheets are shown
in Figs.~\ref{fig:chern}(e-f). While these WCCs
do not vary strongly along $k_z$, they wind
by one unit as they evolve along $k_x$, pumping 
one electron per unit cell from the (0$\bar{1}$0)
to the (010) surface.  The pumped charge is 
restored on each surface as the surface bands 
cross the Fermi level in the bulk energy gap. 

%============================================================
\subsection{TR-invariant trivial insulator}
\label{sec:z2even}
%============================================================

%-------------------------------------------
\begin{figure}[!]
\includegraphics[width=1.\columnwidth]{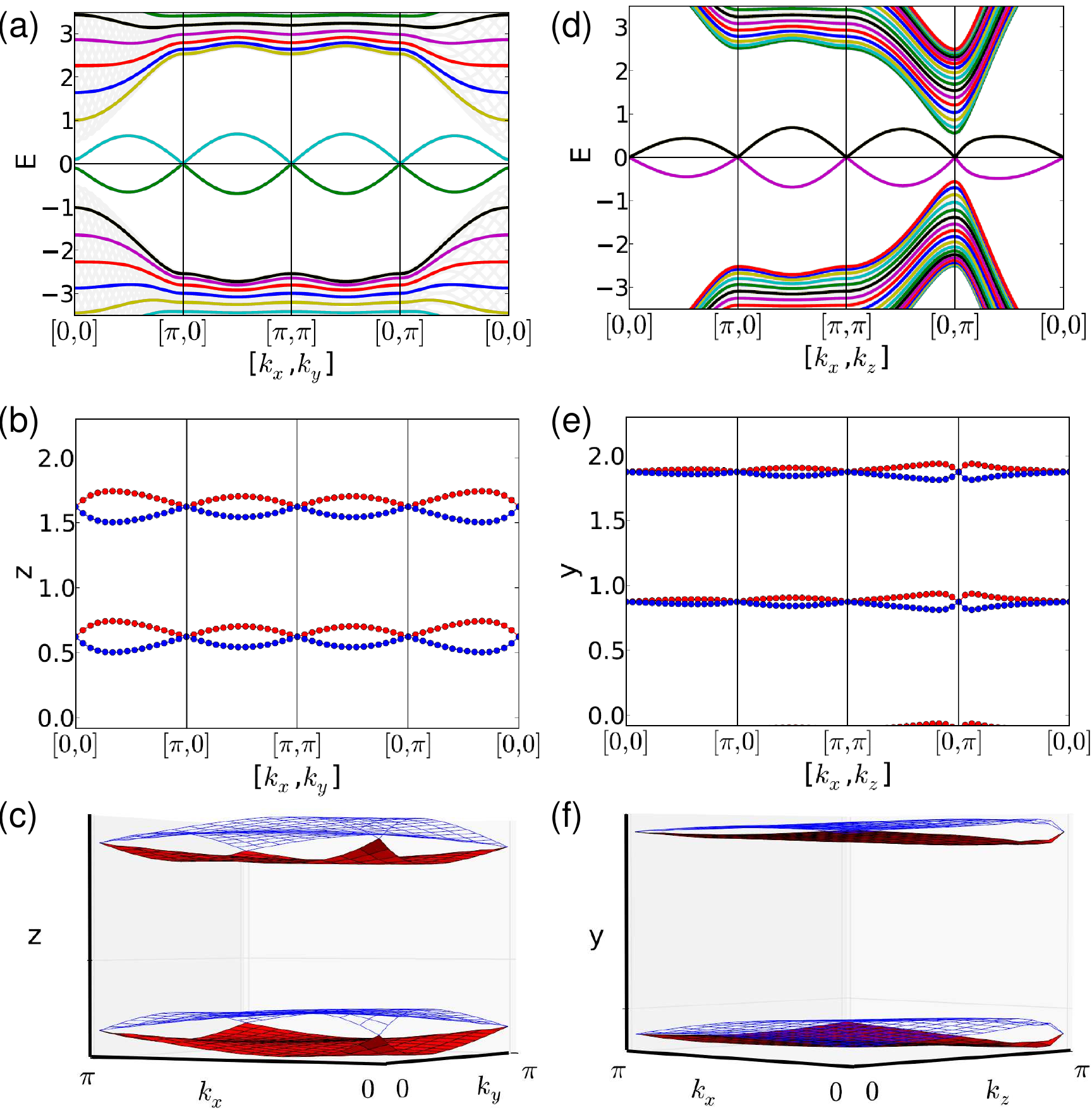}
\caption{\label{fig:km-even}
(Color online)
Surface energy bands (15-layer slab) and WCC sheets
for the TR-invariant FKM
model in the trivial phase ($\alpha\!=\!2.5$).
(a-c) Surface normal and WCCs along $\hat{z}$ vs.\ $(k_x,k_y)$.
(d-f) Surface normal and WCCs along $\hat{y}$ vs.\ $(k_x,k_z)$.
The WCC sheets and surface bands show a trivial behavior
in all directions.}
\end{figure}
%---------------------------------------------

In general, the broken translational symmetry 
at the surface of a band insulator allows for the
existence of surface states in the bulk band gap.
In a topologically trivial insulator, these surface states,
if present, are prone to localization by disorder and can be
removed from the gap by an adiabatic transformation of the
Hamiltonian.  An example of such unprotected surface
states can be seen in
Fig.~\ref{fig:km-even}(a), which shows the
surface states on the
(001) surface of the FKM model in its trivial
insulating phase. The surface bands are
doubly degenerate at the TRI momenta as
required by Kramer's theorem, but nothing
protects them from being adiabatically
pushed to the valence or conduction band.
(The model also happens to have a particle-hole
symmetry which is responsible for the mirror symmetry
along the energy axis, but we do not consider this as
an imposed symmetry here.)
The surface energy bands on the (010) surface,
Fig.~\ref{fig:km-even}(d), show the same trivial behavior,
indicating that this is a topologically trivial insulator.

The trivial topology of this material is equally evident from
the WCC sheets, plotted along $\hat{z}$ and $\hat{y}$ in
Figs.~\ref{fig:km-even}(b-c) and (e-f) respectively.  The WCC
sheets are plotted around the boundary of a quadrant of
the 2D projected BZ in
Figs.~\ref{fig:km-even}(b) and (e), and throughout its interior
in Figs.~\ref{fig:km-even}(c) and (f).  Here there are two WCC
sheets per unit cell (vertical axis) because there are two occupied
energy bands in the four-band model, but the band pairs remain
well separated from their periodic images above and below.
The WCC sheets touch at
the TRI points at the corners of the quarter BZ, as required
by Kramers' theorem, but these Kramers pairs are connected
in all directions in a topologically trivial way.
As a result, the topological index is $\nu_{\mu}=+1$ on all
six TRI faces, signalling a fully trivial topological phase.

%-------------------------------------------
\begin{figure}
    \includegraphics[width=1.\columnwidth]{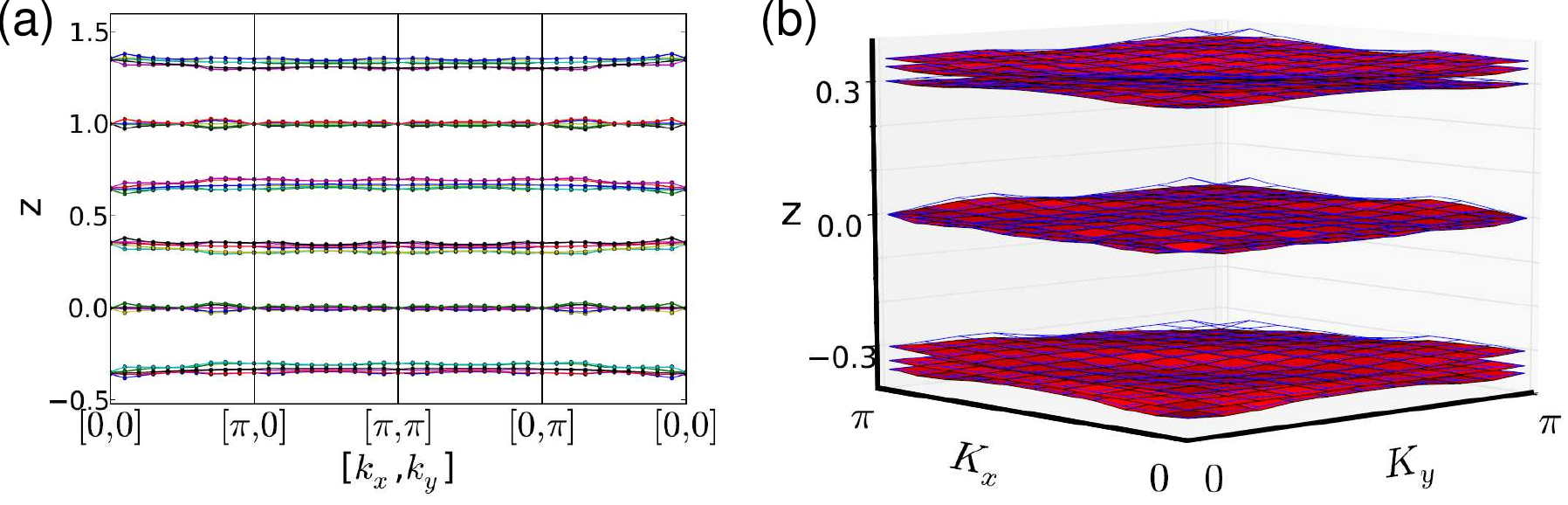}
\caption{\label{fig:sbse}
(Color online)
First-principles WCC sheets along $\hat{z}$ for topologically
trivial Sb$_2$Se$_3$, plotted on (a) the boundary and (b) the interior of the 2D quarter BZ.
The WCCs show trivial behavior as expected.}
\end{figure}
%---------------------------------------------

A similar trivial behavior is seen in the first-principles
WCCs computed for Sb$_2$Se$_3$ as shown in Fig.~\ref{fig:sbse}. The
18 WCC sheets in the quintuple layer come mainly
from the Sb $5p$ and Se $4p$ orbitals.
While having substantial Sb $5p$ character, they are
nevertheless centered on the anion Se sites located
at $z \simeq -0.3c$, $0$, and $0.3c$ in the figure.
While the gap between WCC sheets associated with neighboring quintuple layers, centered at
$0.5c$ in Fig.~\ref{fig:sbse}(a), is not obviously
larger than the other gaps, it nevertheless remains
open across the entire 2D BZ. 
The WCC sheets plotted along the $x$ and
$y$ directions (not shown) display a similar trivial behavior.
Thus, we can conclude that this is a fully trivial insulator,
without having to carry out any surface-state calculation.

%============================================================
\subsection{TR-invariant weak topological insulator}
\label{sec:wz2odd}
%============================================================

\begin{figure}
    \includegraphics[width=\columnwidth]{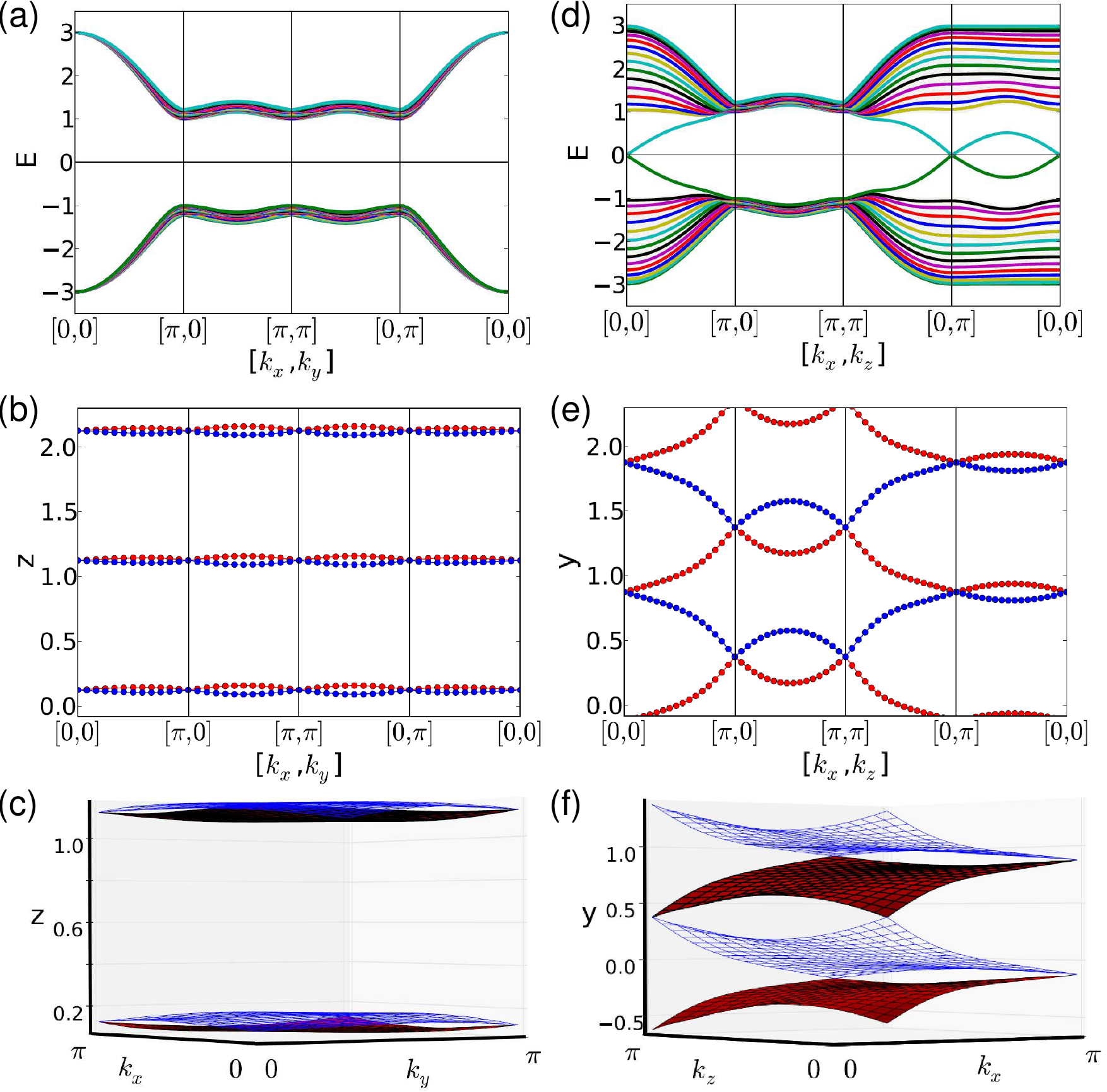}
\caption{\label{fig:km-weak-odd}
(Color online)
Surface energy bands (15-layer slab) and WCC sheets for
the TR-invariant FKM
model in the weak topological phase ($\alpha\!=\!-1$).
(a-c) Surface normal and WCCs along $\hat{z}$ vs.\ $(k_x,k_y)$.
(d-f) Surface normal and WCCs along $\hat{y}$ vs.\ $(k_x,k_z)$.
Only the $(k_x,k_y)$ TRI faces at $k_z\!=\!0$ and $k_z\!=\!\pi$
are $Z_2$-odd.}
\end{figure}
%---------------------------------------------

The FKM model with $-2\leq\alpha\leq0$ is a weak $Z_2$-odd insulator,
as illustrated by our results for $\alpha=-1$ in
Fig.~\ref{fig:km-weak-odd}.  In this case, the crystal can be
thought of as a series of 2D spin-Hall insulators stacked along
the $z$ direction, i.e., the direction of the weakest bond.
Thus, a slab of the model cut normal
to this direction shows no robust surface states
in the bulk energy gap, as shown in
Fig.~\ref{fig:km-weak-odd}(a),
and the WCC sheets along this direction
pair up as they do in a trivial insulator, as
can be seen in Figs.~\ref{fig:km-weak-odd}(b-c).

On the other hand, a slab of a weak
$Z_2$-odd insulator cut through the 2D
spin-Hall sheets should host an even number of
Dirac cones on each surface. These surface
states are shown for an (010) slab of the same FKM
model in Fig.~\ref{fig:km-weak-odd}(d), where
the Dirac cones are visible at $(k_x,k_z)=(0,0)$ and $(0,\pi)$.
These surface bands have a gap-crossing $Z_2$-odd behavior vs.~$k_x$
but not vs.~$k_z$, suggesting that the $(k_x,k_y)$ TRI faces of
the 3D BZ are $Z_2$-odd at $k_z\!=\!0$ and $\pi$, while those on
the $(k_y,k_z)$ faces are $Z_2$-even at $k_x\!=\!0$ and $\pi$.
This is confirmed in Figs.~\ref{fig:km-weak-odd}(e-f), where
the WCC sheets are seen to swap partners vs.~$k_x$ but not
vs.~$k_z$.

The $Z_2$ topological invariants $\nu_\mu$ follow straightforwardly
from the above considerations.  The invariants are
$+1$ for the TRI faces at $k_x\!=\!0$ and $\pi$,
$+1$ for the TRI faces at $k_y\!=\!0$ and $\pi$, and
$-1$ for the TRI faces at $k_z\!=\!0$ and $\pi$.
The conventional index set is then
$[\nu_0;\nu_1 \nu_2 \nu_3]=[+;++-]$,
confirming that this is a weak TI ($\nu_0=+1$)
corresponding to spin-Hall layers stacked along $z$ ($\nu_3=-1$).

%-------------------------------------------
\begin{figure}
    \includegraphics[width=\columnwidth]{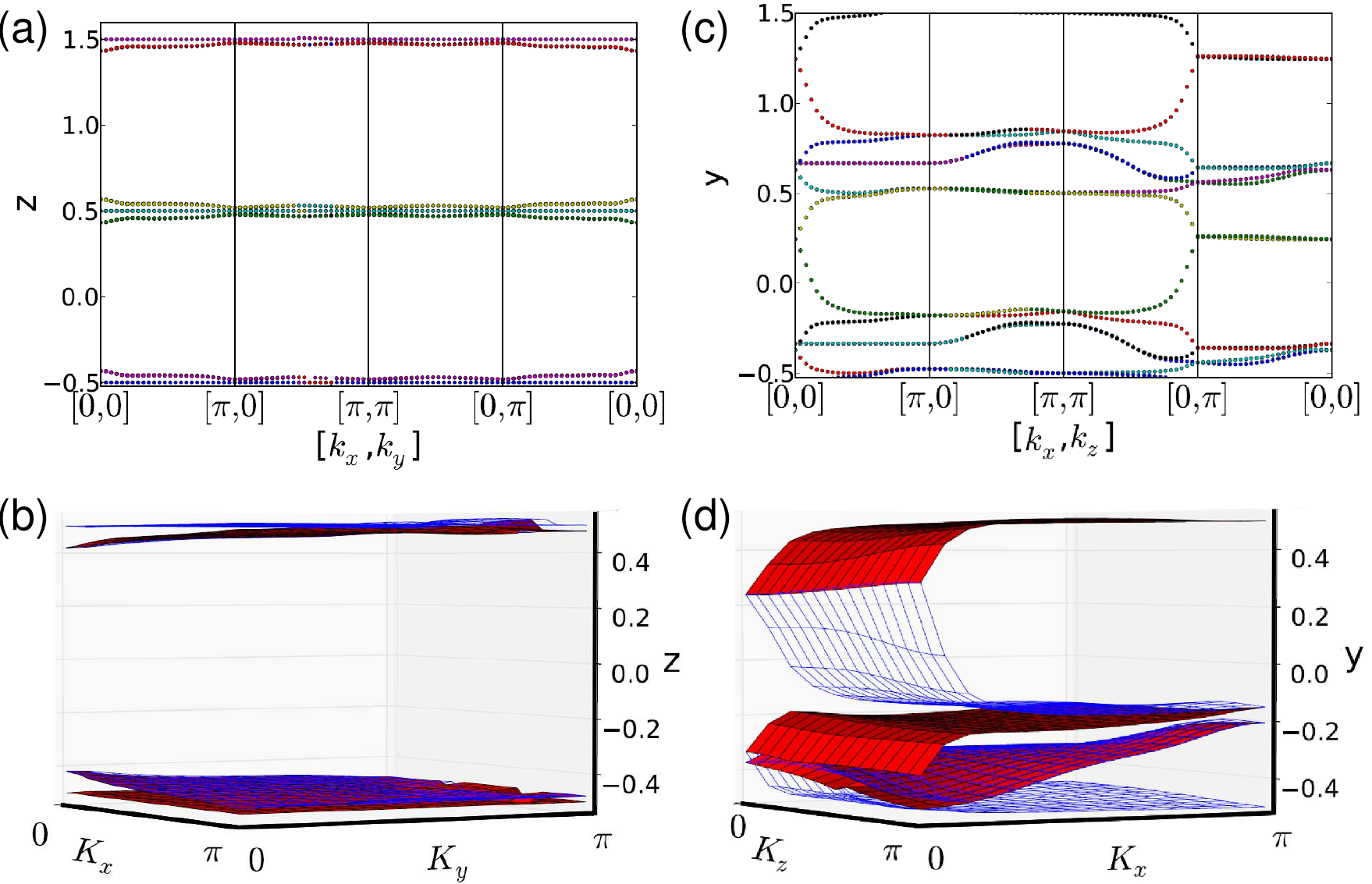}
\caption{\label{fig:3d-khgsb}
(Color online)
First-principles WCC sheets for the weak TI KHgSb.
(a-b) Along $\hat{z}$.
(c-d) Along $\hat{y}$.
Only the $(k_x,k_y)$ TRI faces at $k_z\!=\!0$ and $k_z\!=\!\pi$
are $Z_2$-odd.}
\end{figure}
%---------------------------------------------

We see the same kind of weak topological behavior in our
first-principles calculations of the WCC sheets for KHgSb shown
in Fig.~\ref{fig:3d-khgsb}.
As explained in Sec.~\ref{sec:r-material},
this material is composed of honeycomb HgSb layers that behave
as 2D spin-Hall insulators, stacked along the $z$ direction,
and separated by hexagonal layers of K stuffing atoms.
The pictures look more complicated because there are now
six occupied bands per cell, and thus six WCCs per lattice
constant, and some of the artificial symmetries of the FKM model
are now absent.  However, the topological behavior is similar to
that of Fig.~\ref{fig:km-weak-odd}.
The weak coupling between the HgSb layers
is reflected in the trivial behavior of the
WCC sheets along the (001) direction, Figs.~\ref{fig:3d-khgsb}(a-b),
but plotting the WCCs in a direction cutting across the
honeycomb HgSb layers reveals the topological behavior, as seen in Figs.~\ref{fig:3d-khgsb}(c-d). These WCC
sheets change partners on the  $(k_x,k_y)$ TRI faces at both
$k_z\!=\!0$ and $\pi$,
indicating $\nu_3=-1$ and $\nu_0=+1$, giving the same
$[+,++-]$ set of indices as for the FKM model in its
weak topological phase.  These results are entirely consistent
with the existence of Dirac cones at the $\overline{\Gamma}$ and
$\overline{Z}$ points in the surface bands of an (010) slab
as shown in Ref.~\onlinecite{yan-prl12}.  However, we again
emphasize the convenience of our approach, in which only
primitive-cell bulk calculations are needed.

%============================================================
\subsection{TR-invariant strong topological insulator}
\label{sec:sz2odd}
%============================================================

%-------------------------------------------
\begin{figure}[t]
    \includegraphics[width=\columnwidth]{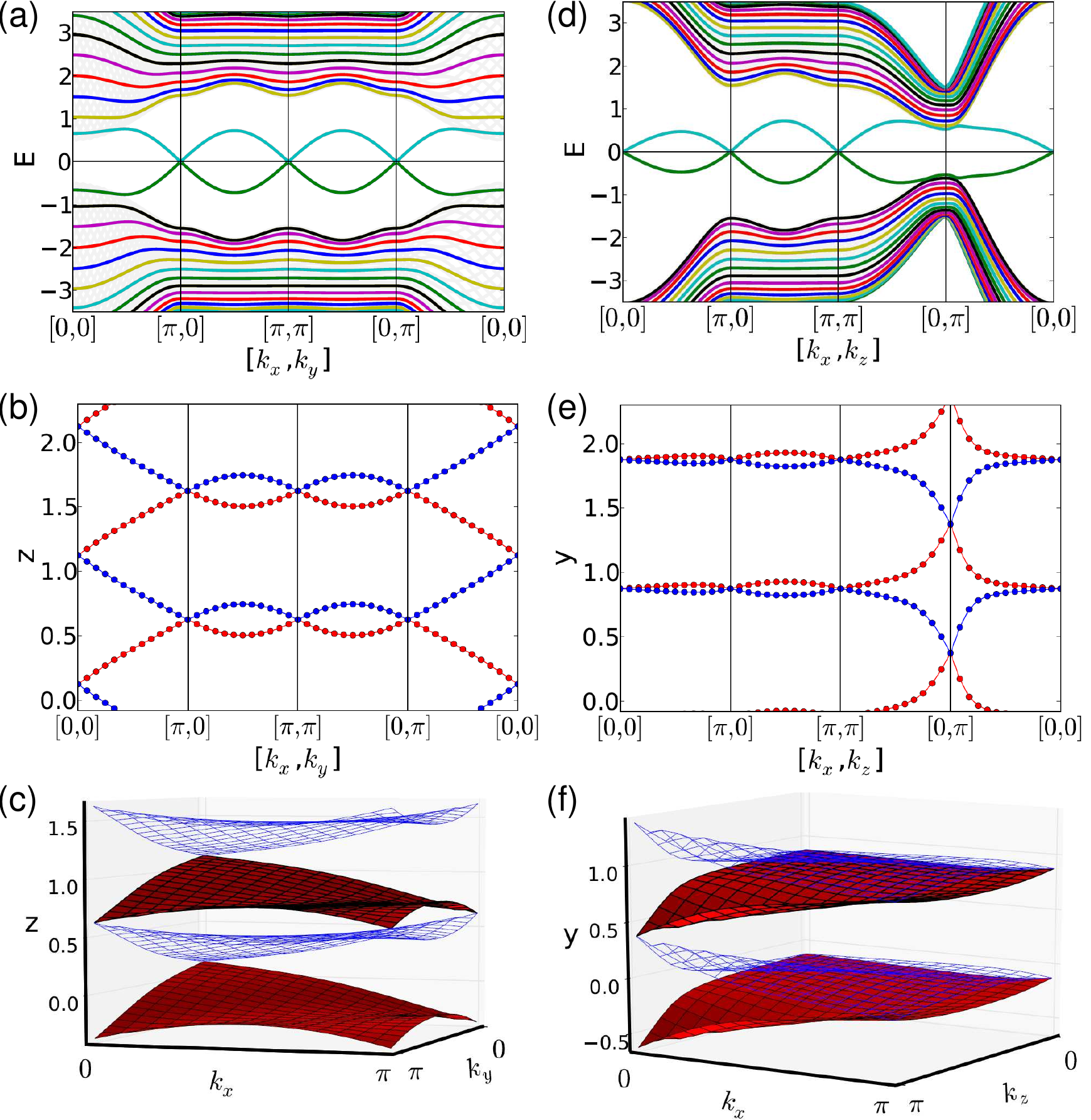}
    \label{fig:wcc-km-strong}
\caption{\label{fig:km-strong}
(Color online)
Surface energy bands (15-layer slab) and WCC sheets for
the TR-invariant FKM
model in the strong topological phase $(\alpha\!=\!1)$.
(a-c) Surface normal and WCCs along $\hat{z}$ vs.\ $(k_x,k_y)$.
(d-f) Surface normal and WCCs along $\hat{y}$ vs.\ $(k_x,k_z)$.
The $(k_x,k_y)$ TRI face at $k_z\!=\!\pi$,
the $(k_x,k_z)$ TRI face at $k_y\!=\!0$, and
the $(k_y,k_z)$ TRI face at $k_x\!=\!0$
are $Z_2$-odd.}
\end{figure}
%---------------------------------------------

In contrast to weak TIs,
the non-trivial behavior of the WCC sheets in
strong $Z_2$ insulators should be evident no matter
what direction is chosen to construct them; there would
be switching of partners for one of the TRI faces
in any chosen direction.  This behavior is illustrated
in  Fig.~\ref{fig:km-strong}, where the surface bands
and WCC sheets are presented for the FKM model in the strong
$Z_2$-odd phase at $\alpha=1$.  Both the surface bands and
the WCC sheets swap partners in
the $(k_x,k_y)$ plane at $k_z\!=\!\pi$,
the $(k_x,k_z)$ plane at $k_y\!=\!0$, and
the $(k_y,k_z)$ plane at $k_x\!=\!0$,
but not on the other three TRI faces.  The set of
topological indices is therefore
$[\nu_0;\nu_1 \nu_2 \nu_3]=[-;++-]$, and the system
is a strong TI.  This is also consistent with the existence of an
odd number of Dirac cones on any surface of a strong $Z_2$
insulator, as is evident in Figs.~\ref{fig:km-strong}(a) and (d),
where three Dirac cones are visible in each case.

\begin{figure}
    \includegraphics[width=\columnwidth]{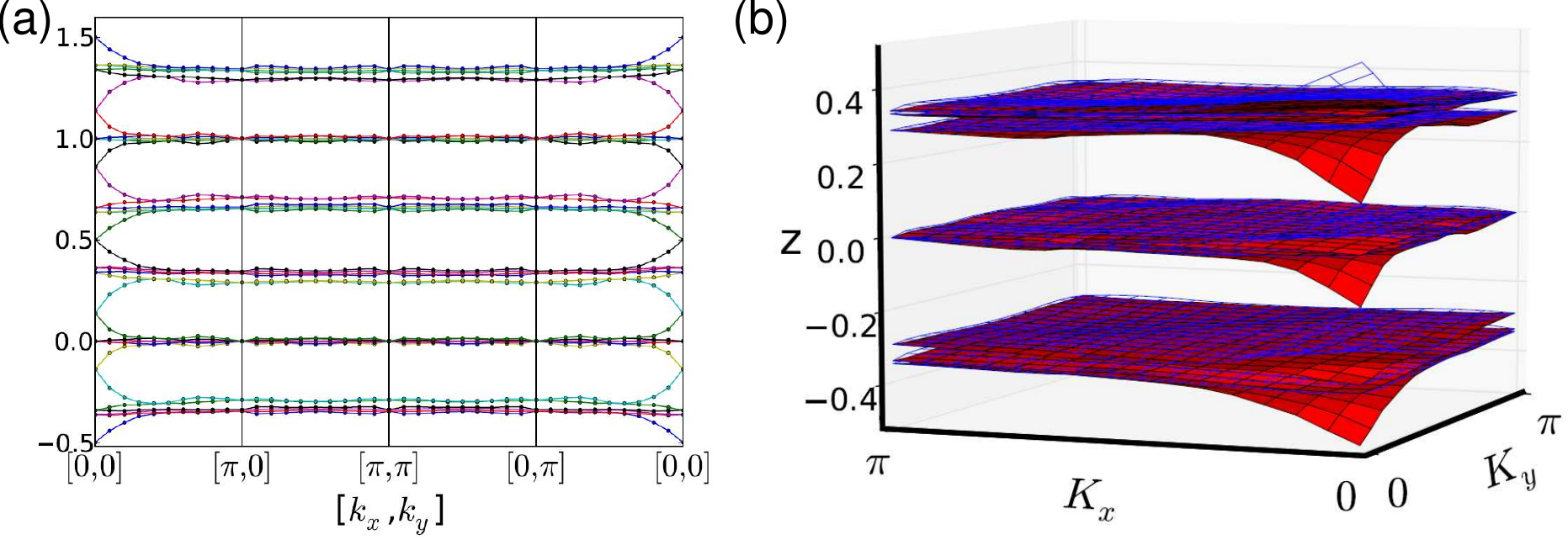}
\caption{\label{fig:bise}
(Color online)
First-principles WCC sheets for
the strong TI Bi$_2$Se$_3$,
plotted on (a) the boundary and (b) the
interior of the 2D quarter BZ.
The WCC sheets on parallel TRI faces (e.g.,
at $k_x\!=\!0$ and $k_x\!=\!\pi$) show
opposite topological behavior.}
\end{figure}

We again confirm that our approach works in the
first-principles context by presenting the WCC sheets along
the $z$ direction (rhombohedral-axis) in
the strong TI Bi$_2$Se$_3$, as shown in Fig.~\ref{fig:bise}.
There are now 18 WCC sheets per cell; in most of the 2D
projected BZ these are clustered in groups of six, with each
of the three clusters located close to the $z$ position
of a layer of Se atoms within the QL.  This is reasonable,
as the Bi and Se atoms can be regarded as cations and anions
respectively, and it is natural to find the Wannier centers
on the anions.  However, this behavior changes drastically
near $\overline{\Gamma}$, where two of the six WCC
sheets in each cluster split off and form a Dirac point
at $\overline{\Gamma}$, signaling the strong TI nature of this
material.  Clearly this results from the band inversion near
$\Gamma$ in the 3D bulk BZ, and is consistent with the existence
of a single Dirac cone at $\overline{\Gamma}$ on the
surface of Bi$_2$Se$_3$, as has been amply demonstrated
by angle-resolved photoemission and other experimental probes.
\cite{hassan-rmp10}
We can again read off the topological indices by noting that
the WCC sheets swap partners in
the $(k_x,k_y)$ plane at $k_z\!=\!0$,
the $(k_x,k_z)$ plane at $k_y\!=\!0$, and
the $(k_y,k_z)$ plane at $k_x\!=\!0$,
but not on the other three TRI faces, so that
$[\nu_0;\nu_1 \nu_2 \nu_3]=[-;+++]$.

%-------------------------------------------
%---------------------------------------------

%============================================================
\subsection{Crystalline topological insulator}
\label{sec:crystalline}
%============================================================

\begin{figure}
   \includegraphics[width=\columnwidth]{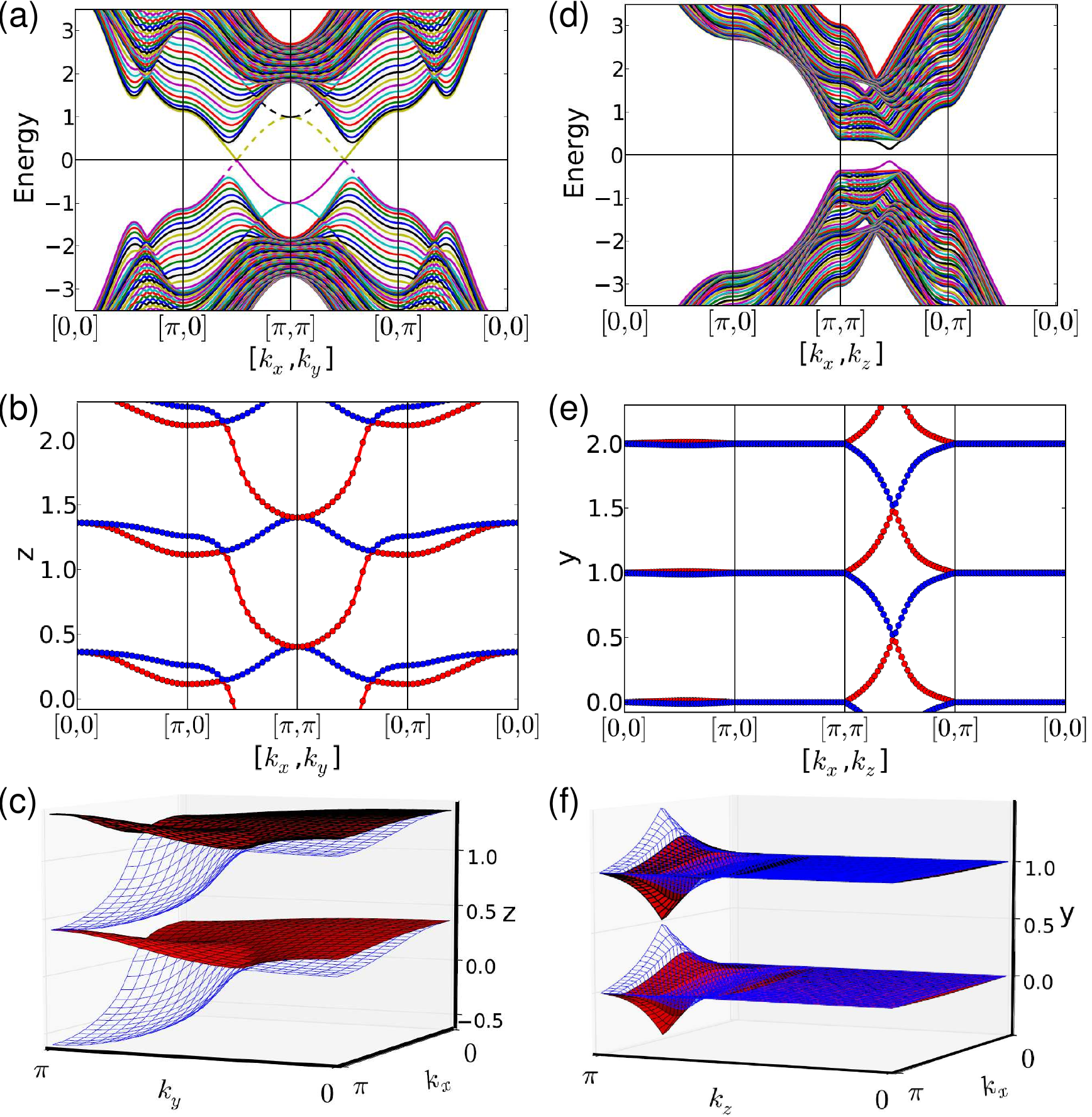}
\caption{\label{fig:fu}
(Color online)
Surface energy bands (24-layer slab) and WCC sheets for
tight-binding model of a crystalline TI.
(a-c) Surface normal and WCCs along $\hat{z}$ vs.\ $(k_x,k_y)$.
(d-f) Surface normal and WCCs along $\hat{y}$ vs.\ $(k_x,k_z)$.
Dashed and solid surface states in (a) reside on the top and
bottom of the (001) slab respectively.
Quadratic band touching and cross-linking in panels (a-c)
signals the crystalline topological phase.}
\end{figure}

In constrast to the systems studied above, Fu's
tetragonal model for a crystalline
TI\cite{fu-prl11} is spinless,
because the non-trivial topology of a
topological crystalline insulator has its
roots in the crystal symmetries
rather than in TR symmetry and spin-orbit
interaction. The TR symmetry in
this scalar model does not guarantee double
degeneracy at the TRI momenta, but its
combination with the crystal $C_4$ symmetry
leads to a two-fold degeneracy of the surface energy bands
at $\overline{\Gamma}=(0,0)$ and at $\overline{\rm M}=(\pi/a,\pi/a)$
for the (001) surface, where
$z$ is chosen along the tetragonal
axis. These surface bands can be seen in
Fig.~\ref{fig:fu}(a) for an (001) slab of
the model. The dashed and solid lines show
the surface states on the two surfaces of
the slab. These bands traverse the energy
gap in a zig-zag manner, and their protected
degeneracy at the $\overline{M}$ point guarantees 
a robust metallic (001) surface.
This non-trivial
behavior is clearer when considering the behavior
of the WCC sheets along the $z$ direction, plotted in 
Figs.~\ref{fig:fu}(b-c). Over most of the 2D BZ, the
$z$ location of these sheets is midway between the
A and B atoms.  The sheets touch two-by-two at
$\overline{\Gamma}$, but they open up
and switch partners on approaching the $\overline{\rm M}$ point. Thus, the WCC undergo the same kind of
switching, and so reflect the same topological
properties, as in the surface energy bands.  
Even the quadratic dispersion of the surface 
bands around $\overline{\rm M}$
is reflected in the WCCs.

%-------------------------------------------
%---------------------------------------------

The $C_4$ symmetry is broken on any surface
other than the (001) surface, which means no
robust surface states are expected on these other
surfaces. Fig.~\ref{fig:fu}(d) confirms this for the case of
an (010) slab of the model.
The energy bands approach each other near a point midway
between $(k_x,k_z)=(\pi,\pi)$ and $(0,\pi)$, but they do not
touch. The WCC sheets show a similar behavior in
Figs.~\ref{fig:fu}(e-f), remaining
trivial except along the segment at $k_z\!=\!\pi$; while
there is a non-avoided crossing along this line, this
appears to be an artifact of some special symmetries of the
model, and is not relevant to the discussion at
hand.\footnote{The topological index for the path from $(\pi,\pi)$
  to $(0,0)$ (third and fourth panels) in Figs.~\ref{fig:fu}(e)
  is even, because a horizontal segment drawn at any chosen $y$
  crosses the sheets an even number of times along this path;
  this is true regardless of whether the crossing is avoided or not.}
Thus, both the surface bands and WCC sheets are
consistent with the trivial topology of an (010) slab of the model.

%============================================================
\section{ SUMMARY }
\label{sec:sum}
%============================================================

In this manuscript, we have explained how the hybrid Wannier
charge centers, or WCC sheets, can be calculated using a
parallel-transport approach along a chosen direction
in a 3D insulator and plotted versus the other $k$-space
dimensions. We have shown that these sheets contain the same topological
information as the surface energy bands, and thus
provide an accessible means of deducing the topological
invariants of the insulator from the bulk properties alone. 
We also show that the linear dispersion
of the surface energy bands at Dirac points in $Z_2$ TIs, and
their quadratic behavior at the gap closure
in topological crystalline insulators,
are replicated by the WCCs.
Moreover, the symmetry group of the WCCs in the 2D BZ include all
the symmetry operators of the surface bands.

We have demonstrated the distinct behavior of the WCC sheets in
trivial, Chern, weak, strong, and crystalline
TIs using various tight-binding 
models.  In addition, we have used
first-principles calculations to illustrate the
calculation of the WCC
sheets in $Z_2$-even Sb$_2$Se$_3$, weak $Z_2$-odd KHgSb, 
and strong $Z_2$-odd Bi$_2$Se$_3$, confirming the conclusions
from the tight-binding models. 

Admittedly, the topological invariants of Chern, TR-invariant,
and crystalline TIs can be deduced in
other ways.  For example, for the TR-invariant case, parity
eigenvalues can be used if inversion symmetry is present; if
not, a calculation of 1D Wannier centers on each 2D TRI face is
sufficient.\cite{soluyanov-prb11,yu-prb11} However, the WCC sheets provide
a unifying description that works in all these cases, allows for
a more intuitive comparison of different kinds of TIs, and provides
deeper insight into the origins of the
non-trivial topology.

The evolution of the WCC sheets as the Hamiltonian
is varied through a trivial-to-topological phase transition, or
carried adiabatically around a loop that pumps the Chern-Simons
axion coupling by a quantum,\cite{qi-prb08,essin-prl09}
would be interesting targets for future studies.
Other phases, such as axion insulators\cite{wan-prl12} and
antiferromagnetic TIs,\cite{mong-prb10}
might also be good subjects for investigation with this tool.
Even in zero-gap Weyl semimetals, the WCC sheets will be well-defined
everywhere except at isolated projected Weyl points in the 2D
BZ, and studying their 
distinct topological properties would be interesting.  Finally,
it would be intriguing to explore whether the WCC concept can be
generalized to topological superconductors.
Thus, we are hopeful that the construction and inspection
of the Wannier charge center sheets will prove to
be a useful tool for the characterization of topological
matter in general.

%============================================================
\acknowledgments
%============================================================
This work was supported by NSF Grant DMR-10-05838.

%%%%%%%%%%%%%%%%
% BIBLIOGRAPHY %
%%%%%%%%%%%%%%%%
\bibliography{p-wcc}
%===============================================================
\end{document}